# Symmetry-controlled orbital Hall effect in IrO$_2$

Michael Patton[1], Dongwook Go[6], Daniel A. Pharis[3], Xiaoxi Huang[3], Gautam Gurung[8,9], Evgeny Y. Tsymbal[4], Daniel C. Ralph[3,7], Mark S. Rzchowski[2], Yuriy Mokrousov[5,6], Chang-Beom Eom[1*]

[1]Department of Materials Science and Engineering, University of Wisconsin-Madison, Madison, Wisconsin 53706, United States.
[2]Department of Physics, University of Wisconsin-Madison, Madison, Wisconsin 53706, United States.
[3]Cornell University, Ithaca, New York 14853, United States.
[4]Department of Physics and Astronomy & Nebraska Center for Materials and Nanoscience, University of Nebraska, Lincoln, NE 68588, United States.
[5]Peter Grünberg Institut, Forschungszentrum Jülich and JARA, Jülich, Germany
[6]Institute of Physics, Johannes Gutenberg University Mainz, Mainz, Germany
[7]Kavli Institute at Cornell for Nanoscale Science, Ithaca, New York 14853, United States.
[8]Clarendon Laboratory, Department of Physics, University of Oxford, Parks Road, Oxford, OX1 3PU UK
[9]Trinity College, University of Oxford, Oxford, OX1 3BH UK
*Corresponding author: ceom@wisc.edu

**Abstract**

Recent discovery of orbital currents in several material platforms including light element metals has opened new possibilities for exploring novel transport phenomena and applications to spin-orbitronic devices. These orbital currents, similar to spin currents, have the ability to generate torque on adjacent magnetic layers, opening a new avenue for efficient spintronic devices. However, separating spin and orbital currents has been one of the major challenges. Here, we show evidence for large conventional as well as unconventional spin and orbital currents in IrO$_2$ and disentangle them by crystal symmetry. We study the anisotropic spin and orbital Hall effects in IrO$_2$ (001), (100), and (111) orientations and find unconventional z-polarized orbital torques using angular spin torque ferromagnetic resonance of IrO$_2$/Ni heterostructures, which are in agreement regarding the relative signs with theoretical calculations of spin and orbital Hall conductivity. This work provides a promising route towards highly efficient low power spintronic and orbitronic devices in oxide heterostructures.

**Introduction**

The recent discovery of orbital currents has opened the possibility of utilizing the orbital degree of freedom of electrons as an information carrier, promoting orbitronics as a promising candidate for next-generation quantum technology beyond the conventional electronics.[1,2] The orbital Hall effect (OHE)[3–8] – a phenomenon where an orbital-polarized current is generated by an external electric field – has been verified in various experiments via magneto-optical effect[9,10], magnetoresistance[11–14], and current-induced torques[15–22]. The OHE is particularly promising for spintronics due to its ability to manipulate magnetization when orbital angular momentum is injected to a ferromagnet[23], which may surpass the efficiency of other conventional mechanisms based on spin currents such as spin Hall effect (SHE). Moreover, the OHE can be much stronger than the SHE in broad range of materials, including environment-friendly and non-toxic elements[24].

However, the main difficulty in detecting orbital current by current-induced torque measurements has been to unambiguously disentangle the spin and orbital current contributions to the torque. Previous works have partly resolved the issue by measuring the current-induced torque for different ferromagnets[16,18], which sensitively depends on the correlation between the spin and orbital angular momenta in the magnet's electronic structure, and also by converting orbital current to spin current by an insertion layer with strong spin-orbit coupling (SOC) such as Pt[15,19]. Contributions from orbital current have furthermore been identified by examining the dependence of current-induced torque on the ferromagnetic layer thickness, which can have a significantly longer length scale than the spin dephasing length ~< 1nm [20,21,25]. However, most of the experiments performed so far rely on several assumptions and quantitative estimation by theory.

In this work, we demonstrate an unambiguous way to disentangle the SHE and OHE by utilizing the crystal-symmetry-dependent orbital Hall conductivity (OHC) and spin Hall conductivity (SHC) tensors. Because the energy bands with different orbital characters are split by the crystal-field potential, the OHC varies significantly depending on the crystal orientation. Here, we use spin torque ferromagnetic resonance (ST-FMR) in various orientations of $IrO_2$ with a Ni detection overlay to uncover the crystallographic dependance of the OHC orbital torques. In comparison with theoretical calculations, we find qualitative agreement for the relative signs for the experimentally determined OHC and SHC in the (001) and (100) orientations. We also find orbital contributions for the unconventional out-of-plane polarization in the (111) orientation. These results can provide alternative approaches for highly efficient low power spintronic devices.

**Results and Discussion**

We study IrO$_2$ epitaxial thin films grown via RF magnetron sputtering on TiO$_2$ (001), (100), and (111) substrates (see experimental section for growth details). We use ST-FMR measurements to determine the SHC and OHC (see Methods and Supplemental Note 2 for more details). To experimentally distinguish the spin and orbital contributions to the torque on the magnetization, we perform separate measurements on IrO$_2$/Py and IrO$_2$/Ni. We assume that Py is insensitive to orbital currents, while Ni is susceptible to both spin and orbital currents[20]. Importantly, we utilize the distinct absorption lengths for the spin and orbital angular momenta in Ni, where the orbital torque is long-ranged over several nanometers [20,25]. By varying the thickness of the Ni layer, we can fit the experimental results to:

$$\sigma^i_{jk\ net} = \sigma^i_{jk\ SHC} + \sigma^i_{jk\ OHC,eff}[1 - \text{sech}\left(\frac{t_{Ni}}{\lambda_{Ni}}\right)] \tag{1}$$

where $\sigma^i_{jk\ SHC}$ is the SHC contribution, $\sigma^i_{jk\ OHC,eff}$ is the effective OHC contribution, and $\lambda_{Ni}$ is the characteristic length for the absorption of orbital currents. It is important to note that $\sigma^i_{jk\ net}$ is proportional to the total torque acting on the local magnetic moment of Ni. It incorporates the two separate contributions from the angular-momentum currents in IrO$_2$ and the efficiency of transmission of those angular momentum currents across the interface to a apply a torque to the magnetic layer.

Table 1: Calculated spin and orbital Hall conductivity for IrO$_2$ in the (001) basis.

| | $\sigma^x$ | $\sigma^y$ | $\sigma^z$ |
|---|---|---|---|
| IrO$_2$<br>a = b = 4.498 Å<br>c = 3.154 Å<br>x∥ [100] y∥ [010] z∥ [001] | $\begin{bmatrix} \sigma^x_{xx} & \sigma^x_{xy} & \sigma^x_{xz} \\ \sigma^x_{yx} & \sigma^x_{yy} & \sigma^x_{yz} \\ \sigma^x_{zx} & \sigma^x_{zy} & \sigma^x_{zz} \end{bmatrix}$ | $\begin{bmatrix} \sigma^y_{xx} & \sigma^y_{xy} & \sigma^y_{xz} \\ \sigma^y_{yx} & \sigma^y_{yy} & \sigma^y_{yz} \\ \sigma^y_{zx} & \sigma^y_{zy} & \sigma^y_{zz} \end{bmatrix}$ | $\begin{bmatrix} \sigma^z_{xx} & \sigma^z_{xy} & \sigma^z_{xz} \\ \sigma^z_{yx} & \sigma^z_{yy} & \sigma^z_{yz} \\ \sigma^z_{zx} & \sigma^z_{zy} & \sigma^z_{zz} \end{bmatrix}$ |
| SHC<br>($\frac{\hbar}{e}$ (Ω cm)$^{-1}$) | $\begin{bmatrix} 0 & 0 & 0 \\ 0 & 0 & 162 \\ 0 & -254 & 0 \end{bmatrix}$ | $\begin{bmatrix} 0 & 0 & -162 \\ 0 & 0 & 0 \\ 254 & 0 & 0 \end{bmatrix}$ | $\begin{bmatrix} 0 & 18 & 0 \\ -18 & 0 & 0 \\ 0 & 0 & 0 \end{bmatrix}$ |
| OHC<br>($\frac{\hbar}{e}$ (Ω cm)$^{-1}$) | $\begin{bmatrix} 0 & 0 & 0 \\ 0 & 0 & 160 \\ 0 & 77 & 0 \end{bmatrix}$ | $\begin{bmatrix} 0 & 0 & -160 \\ 0 & 0 & 0 \\ -77 & 0 & 0 \end{bmatrix}$ | $\begin{bmatrix} 0 & -194 & 0 \\ 194 & 0 & 0 \\ 0 & 0 & 0 \end{bmatrix}$ |

Based on our DFT calculations, IrO$_2$ is predicted to exhibit strong anisotropy of both OHE and SHE depending on the crystal orientation. The SHC(OHC) denoted as $\sigma_{jk}^{i}$, where *i* is the spin(orbital) polarization direction, *j* is the spin flow direction, and *k* is the charge current direction, is a 27-element property tensor of a material. For high symmetry materials only some of these elements are non-zero due to crystal symmetry restrictions. For IrO$_2$, the symmetry of the tetragonal rutile crystal structure only allows for 3 non-zero independent SHC(OHC) terms including $\sigma_{zy}^{x}$, $\sigma_{yz}^{x}$, and $\sigma_{xy}^{z}$. Note that both SHC and OHC are subject to the same symmetry constraints, but their quantitative values may differ significantly depending on the orbital and spin characters of bands. The IrO$_2$ calculation results for both SHC and OHC for IrO$_2$ are shown in Table 1. We use a sign convention such that the conventional SHC of Pt, $\sigma_{zx}^{y}$, is negative. Due to the anisotropy of the SHC and OHC tensors, measuring the contributions from spin polarization and orbital polarization have different dependence on the charge current direction. For example, considering the (001) orientation with charge current along the [$\bar{1}$00] direction, spin polarization along the [010] direction, and spin current along the [001] direction is calculated to generate a positive SHC and negative OHC as shown in Fig. 1a where the orbital Berry curvatures exhibit highly anisotropic feature, and their hotspots can significantly differ in the Brillouin zone. This is the direct consequence of the anisotropic orbital-dependent level splitting. Next, considering the (100) orientation with charge current along the [010] direction, spin polarization along the [00$\bar{1}$] direction, and spin current along the [100] direction should generate a positive SHC and positive OHC as shown in Fig. 1b where the spin and orbital Berry curvatures both show positive hotspots. Lastly, considering the (010) orientation with charge current along the [001] direction, spin polarization along the [$\bar{1}$00] direction, and spin current along the [010] direction should generate a positive SHC and negative OHC as shown in Fig. 1c where the spin Berry curvature shows positive hotspots, and the orbital Berry curvature shows negative hotspots. This demonstrates that the anisotropy of SHC and OHC in IrO$_2$ can be distinguished by measuring $\sigma_{jk\ net}^{i}$ for the different crystal orientations.

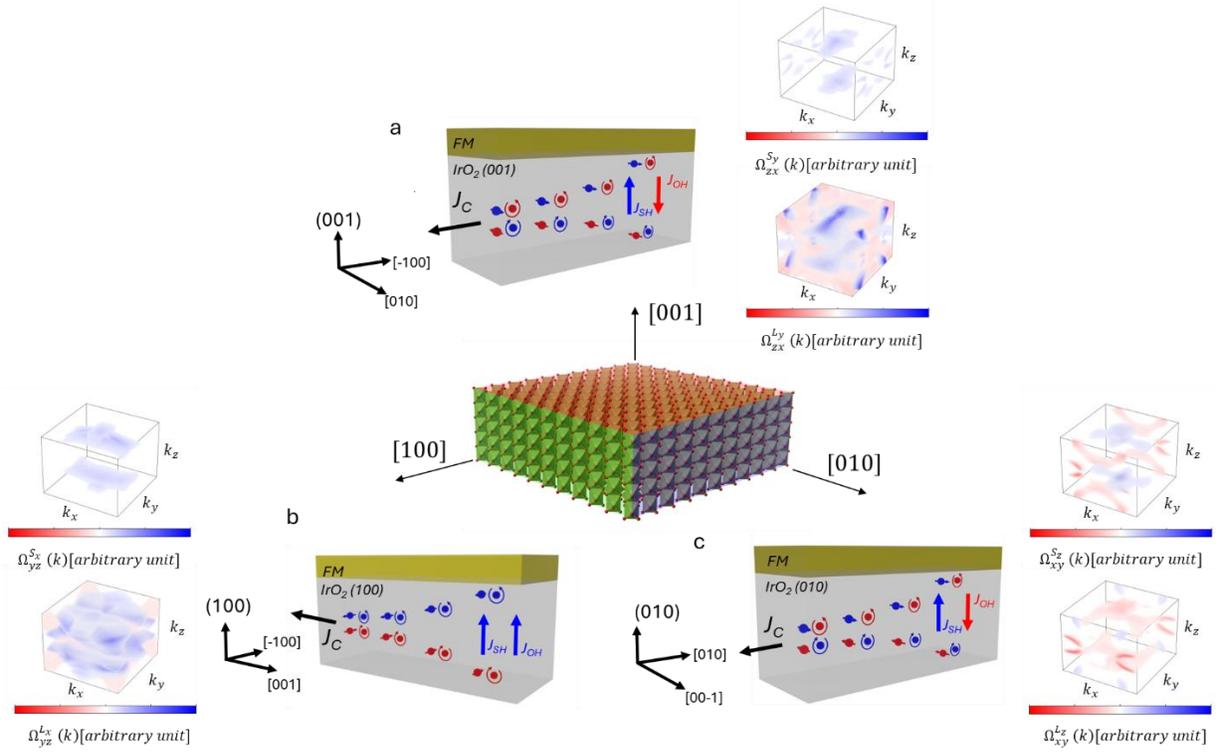

Figure 1: a, spin Hall conductivity and orbital Hall conductivity for IrO$_2$ for a, the (001), b, the (100), and c, the (010) orientations with spin and orbital Berry curvature for each orientation.

To determine the $\sigma_{zy}^x$, $\sigma_{yz}^x$, and $\sigma_{xy}^z$ terms, we measure the (001) orientation which gives $\sigma_{zy}^x$ with charge current along any in-plane direction, and we measure the (100) orientation which gives $-\sigma_{yz}^x$ with charge along the [001] direction, and $-\sigma_{xy}^z$ with charge current along the [010] direction. The experimental results can be seen in Fig. 2 where the value at t$_{Ni}$ = 0 is the reference point from the IrO$_2$/Py bilayers which should have very little sensitivity to OHC contributions, and thus represent the spin current contribution. The results for the (001) orientation, which allows to probe $\sigma_{zy}^x$, show a decreasing trend with increasing Ni thickness, implying a negative orbital Hall contribution and exhibiting an orbital diffusion length of 5.8 ± 1nm. This qualitatively agrees with the relative signs for SHC and OHC from the theoretical calculations. Next, we find OHC contributions with the same sign as SHC in the (100) orientation along the [001] direction with orbital diffusion length of 4 ± 0.6nm and OHC contributions with opposite sign to SHC for charge current along [010] with orbital diffusion length of 2.1 ± 0.7nm, agreeing qualitatively to the theoretical predictions.

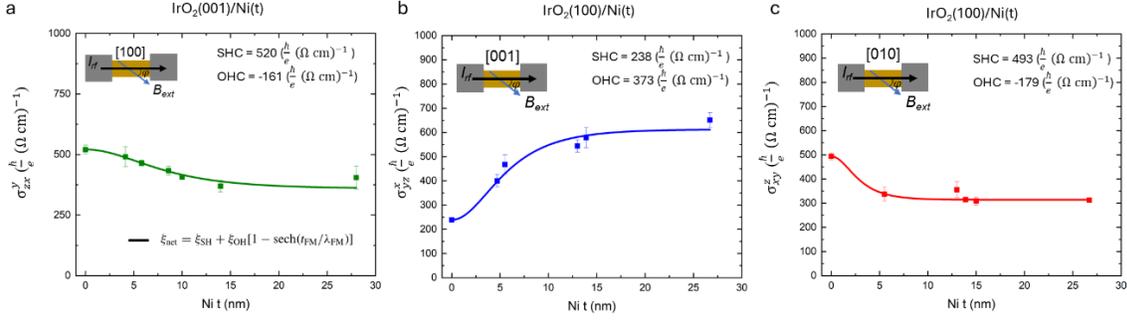

Figure 2: a Net SHC plus OHC contributions vs Ni thickness for (001) oriented IrO$_2$, b for (100) oriented IrO$_2$ with charge current along the [001] direction, and c for (100) oriented IrO$_2$ with charge current along the [010] direction.

Lastly, we study the OHC contributions in the IrO$_2$ (111) orientation, which hosts unconventional SOTs due to the low crystalline symmetry in this orientation. The OHC contributions can be seen in both the [11$\bar{2}$] (defined as capital *Y*) and [$\bar{1}$10] (defined as capital *X*) directions seen in Fig. 3a and b, respectively. In both cases, the SHC and OHC show positive contributions. Additionally, we study the angular ST-FMR with current along the [1$\bar{1}$0] which should generate the unconventional z-spin polarized current. In the IrO$_2$ (111)/Py samples, we found a small non-zero unconventional z-spin SHC of -8 $\pm$ 1 ($\frac{\hbar}{e}$ ($\Omega$ cm)$^{-1}$). In the IrO$_2$ (111)/Ni samples, we find a large increase with increasing thickness for Ni, indicating a large unconventional OHC with z-polarization of -40 $\pm$ 8 ($\frac{\hbar}{e}$ ($\Omega$ cm)$^{-1}$) seen in Fig. 3c.

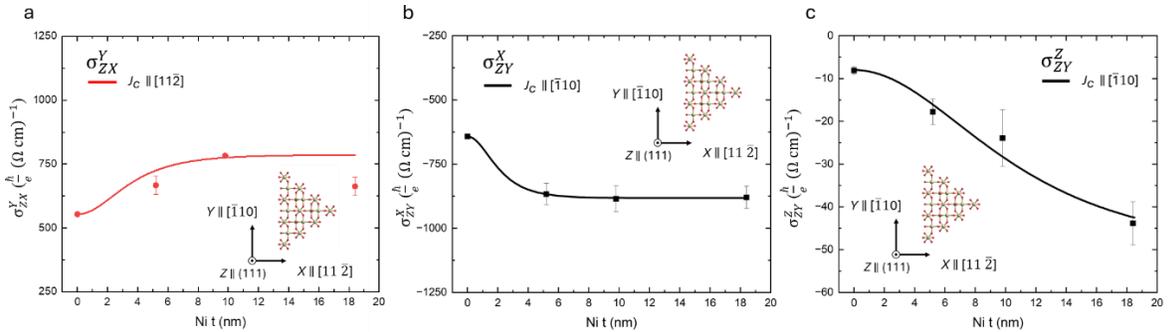

Figure 3: a Net SHC plus OHC contributions for IrO$_2$ (111) for various Ni thickness for charge current along the [11$\bar{2}$], b conventional OHC contributions for IrO$_2$ (111) for various Ni thickness for charge current along the [$\bar{1}$10], and c, the unconventional z-polarized net SHC plus OHC contributions for charge current along the [$\bar{1}$10]. Here, we define X along [$\bar{1}$10], Y along [11$\bar{2}$], and Z perpendicular to the (111) surface.

As an independent check, we have also measured the orbital-spin conversion through a Pt conversion layer, as a second approach to probe the orbital Hall effect[17,20,21]. We use DC-tuned ST-FMR as an alternative approach to probe the effective spin Hall angle Supplemental Note 3 shows a decrease in the effective spin Hall angle for the (001) oriented $IrO_2$/Pt/Py/Pt sample, indicating a negative orbital Hall effect contribution. This is in agreement with both the $IrO_2$/Ni results and theoretical calculations. We also find an increase in the effective spin Hall angle for the (100) oriented $IrO_2$/Pt/Py/Pt sample with charge current along the [001] direction, in agreement with both the $IrO_2$/Ni results and theoretical calculations. Lastly, we find an increase in the effective spin Hall angle magnitude for the (100) oriented $IrO_2$/Pt/Py/Pt sample with charge current along the [010] direction while showing a negative spin Hall angle. Qualitatively, the large negative increase in the spin Hall angle, which indicates a large negative orbital Hall effect contribution, is consistent with both the $IrO_2$/Ni results and theoretical calculations. However, the small negative spin Hall angle in the $IrO_2$(001)/Py control sample found using DC-tuned ST-FMR is inconsistent with the angular ST-FMR results for both magnitude and sign which could be due to known inconsistencies using the DC-tuned ST-FMR technique[26]. Nevertheless, these results qualitatively support the main findings outlined in the main text showing consistent trends in the sign of the orbital Hall effect in $IrO_2$.

**Conclusion**

These results demonstrate a large orbital Hall conductivity present in $IrO_2$ with a strong dependence on crystallographic symmetry. We extract the OHC from the experimental torque measurements by measuring samples with Ni thicknesses spanning the absorption length of orbital angular momentum for the (001), (100), and (111) oriented $IrO_2$. In the (001) orientation, a decreasing experimental torque signal with increasing Ni thickness indicates an OHC contribution opposite in sign to the SHC. For the (100) orientation, we find similar OHC contributions for current along the [010] and [00$\bar{1}$] directions, agreeing with theoretical predictions. We also find evidence for a large unconventional OHC in $IrO_2$ (111). These results provide design approaches for efficient spintronic and orbitronic devices by tuning the strength of the SHE versus OHE via the crystallographic orientation. Additionally, these results offer an alternative approach for highly efficient field-free magnetic switching using unconventional out-of-plane spin and orbital currents. Further experiments using PMA switching can give insight into the switching efficiency for orbital currents compared to spin currents by using $IrO_2$ (111)/Pt bilayers. This work also opens new avenues for orbitronics using oxide systems which have additional tuning knobs such as octahedral rotations, strain engineering, stoichiometry, and other thin film engineering tools, which are highly sensitive to orbital-dependent energy splitting and strongly affects transport and dynamic properties of orbital angular momentum. Using these engineering approaches in other oxide systems that may host large OHC could demonstrate the advantages of using oxide systems for spintronic and orbitronic applications.

**Author Contributions:** M.P., D.G., D.R., Y.M., and C.B.E. conceived the project. M.P. carried out the thin film growth, device fabrication, and spintronic measurements. D.P. and X.H. carried out additional device fabrication and supporting spintronic measurements. D.G. and G.G. performed theoretical calculations. C.B.E., M.S.R., D.C.R., Y.M., and E.Y.T. supervised the study. All authors discussed the results and commented on the manuscript. C.B.E. directed the research.

**Acknowledgments**

CBE acknowledges support for this research through a Vannevar Bush Faculty Fellowship (ONR N00014-20-1-2844), the Gordon and Betty Moore Foundation's EPiQS Initiative, Grant GBMF9065. Transport measurement at the University of Wisconsin–Madison was supported by the US Department of Energy (DOE), Office of Science, Office of Basic Energy Sciences (BES), under award number DE-FG02-06ER46327. D.G. and Y.M. were supported by the EIC Pathfinder OPEN grant 101129641 "OBELIX. Measurements at Cornell were supported by the DOE under award number DE-SC0017671. The work at UNL was partly supported by the National Science Foundation through the EPSCoR RII Track-1 program (Grant OIA-2044049).

**Methods**

*Sample growth, fabrication, and characterization.*

Epitaxial $IrO_2$ was grown on $TiO_2$ (001), (100), and (111) substrates by RF magnetron sputtering followed by *in situ* growth of ferromagnetic permalloy $Ni_{81}Fe_{19}$ (Py). The $IrO_2$ films were grown at 320°C at a pressure of 20 mTorr with 10% oxygen partial pressure. The target power was 20 W. After growth the sample was cooled in an $O_2$ atmosphere. Ni was then grown in situ at room temperature, 8 mTorr of Ar, power of 35W, and a background pressure of 3E-7 Torr. For $IrO_2$/Py samples, Py was grown at 35W at 3.5mTorr. The samples were then fabricated using photolithography and ion beam milling, followed by sputter deposition of 100 nm Pt/10nm Ti and lift off techniques for the electrodes.

*ST-FMR measurements*

During the ST-FMR measurements, a microwave current was applied at a fixed frequency (3-12GHz) and fixed power (10-13 dBm) while sweeping an in-plane magnetic field through the Py (or Ni) resonance conditions from 0 to 0.25 T. The microwave current was modulated at a fixed frequency of 437 Hz and the mixing voltage across the device was measured using a lock-in amplifier. The mixing voltage was fitted vs applied field to extract the symmetric and antisymmetric Lorentzian components. For the angular-dependent ST-FMR, the applied field was rotated in-plane 360° and the symmetric and antisymmetric components were plotted as a function of angle. The out-of-plane ($\tau_\perp$) and the in-plane ($\tau_\parallel$) torques are proportional to the mixing voltage $V_{mix}$ as the ferromagnetic layer goes through its resonance condition which can be fitted as a sum of a symmetric and an antisymmetric Lorentzian:

$$V_{mix,S} = -\frac{I_{rf}}{2}\left(\frac{dR}{d\varphi}\right)\frac{1}{\alpha(2\mu_0 H_{FMR}+\mu_0 M_{eff})}\tau_\parallel$$

$$V_{mix,A} = -\frac{I_{rf}}{2}\left(\frac{dR}{d\varphi}\right)\frac{\sqrt{1+M_{eff}/H_{FMR}}}{\alpha(2\mu_0 H_{FMR}+\mu_0 M_{eff})}\tau_\perp.$$

Here $I_{rf}$ is the RF current calibrated using Joule heating experiments, $R$ is the resistance of the device, $\varphi$ is the magnetization angle with respect to the applied current, $\alpha$ is the Gilbert damping coefficient, $\mu_0 H_{FMR}$ is the resonance field, and $\mu_0 M_{eff}$ is the effective magnetization. The effective magnetization of Py or Ni can be obtained using Kittel's equation $f = \frac{\gamma}{2\pi}\sqrt{(H_{FMR}+H_K)(H_{FMR}+H_K+M_{eff})}$ where $\gamma$ is the gyromagnetic ratio and $H_K$ is the in-plane anisotropy field. For $IrO_2$/Py samples we find $M_{eff}$ to range between 0.6-0.85 T and for $IrO_2$/Ni samples we find $M_{eff}$ to range between 0.2-0.3 T. Additional unconventional torques that have different spin polarization direction contribute to the angular dependance resulting in a more general form of the angular dependance:

$$\tau_\parallel = \tau_{x,AD}\sin(\varphi) + \tau_{y,AD}\cos(\varphi) + \tau_{z,FL}$$

$$\tau_\perp = \tau_{x,FL}\sin(\varphi) + \tau_{y,FL}\cos(\varphi) + \tau_{z,AD}$$

$V_{mix,S}$ and $V_{mix,A}$ can then be expressed in the form of $\sin(2\varphi)(\tau_{x,AD}\sin(\varphi) + \tau_{y,AD}\cos(\varphi) + \tau_{z,FL})$ and $\sin(2\varphi)(\tau_{x,FL}\sin(\varphi) + \tau_{y,FL}\cos(\varphi) + \tau_{z,AD})$, respectively. The spin Hall conductivity $\sigma_{jk}^i$ can then be determined using $\sigma_{jk}^i = \frac{\theta_i}{\rho_{IrO2}}\frac{\hbar}{2e}$, where $\rho_{IrO2}$ is the resistivity and $\theta_i$ is defined as:

$$\theta_i = \tau_{i,AD} \frac{2e\mu_0 M_s t_{FM}}{\gamma \hbar J}$$

Where $t_{FM}$ is the thickness of the ferromagnetic layer (Py or Ni), $M_s$ is the saturation magnetization per unit volume, and $J$ is the charge current density in the IrO$_2$. The resistivities were determined using Van der Pauw measurements and found to be 105 $\mu\Omega cm$ for (001), 226 $\mu\Omega cm$ for (100) along the [010] direction, 160 $\mu\Omega cm$ for (100) along the [001] direction, 95 $\mu\Omega cm$ for (111) along the [11-2] direction, and 105 $\mu\Omega cm$ for (111) along the [1-10] direction.

*First-principles calculation*

For the computation of the OHC and SHC in IrO$_2$, we carry out the three-step calculation based on first-principles methods. In the first step, we perform self-consistent density functional theory (DFT) calculation, by employing the FLEUR code[27] which implements the full-potential linearly augmented plane wave (FLAPW) method[28]. For the exchange and correlation effects, we use the Perdew-Burke-Ernzerhof functional[29] based on the generalized gradient approximation. The lattice constants are set $a_1 = a_2 = 8.50 a_0$ and $a_3 = 5.96 a_0$ in the rutile structure, where $a_0$ is the Bohr radius. The fractional coordinates of atoms are given by

| Atom | $c_1$ | $c_2$ | $c_3$ |
| --- | --- | --- | --- |
| Ir-1 | 0.0 | 0.0 | 0.0 |
| Ir-2 | 0.5 | 0.5 | 0.5 |
| O-1 | 0.1916 | 0.8084 | 0.5 |
| O-2 | 0.3084 | 0.3084 | 0.0 |
| O-3 | 0.6916 | 0.6916 | 0.0 |
| O-4 | 0.8084 | 0.1916 | 0.0 |

by which the relative position of an atom with respect to the center of the unit cell is $\delta \boldsymbol{r} = c_1 \boldsymbol{a}_1 + c_2 \boldsymbol{a}_2 + c_3 \boldsymbol{a}_3$. The following calculation parameters specific to the FLAPW method are used: $R_{\text{Ir}} = 2.30 a_0$ and $R_{\text{O}} = 1.30 a_0$ for the muffin-tin ardii of Ir and O atoms, respectively, $l_{\max} = 12$ for the harmonic expansion in the muffin-tin for both Ir and O atoms, and $K_{\max} = 5.0 a_0^{-1}$ for the plane wave cutoff in the interstitial region. We sample the **k**-points in the first Brillouin zone on the $12 \times 12 \times 16$ Monkhorst-Pack mesh.

In the second step, from the converged electronic structure from the DFT calculation, i.e. the potential and Kohn-Sham states, we construct maximally localized Wannier functions (MLWFs) by employ the WANNIER90 code[30], which is interfaced with the FLEUR code[30]. We project the Kohn-Sham states onto $d_{xy}, d_{yz}, d_{zx}, d_{x^2-y^2}, d_{z^2}$ orbitals on Ir site and $p_x, p_y, p_z$ orbitals on O site as the initial guess and iteratively minimize the spread of the Wannier functions. For the disentanglement, we set the maximum of the frozen energy window 3 eV above the Fermi energy. All the necessary operators (Hamiltonian, position, orbital angular momentum, and spin), which are first represented in the Kohn-Sham states in **k**-space, are transformed into the representations based on the MLWFs in real space. We confirm the consistency between the Wannier and FLAPW methods by comparing their band structures (Supplemental Note 4, Fig. S7).

In the third step, we diagonalize the Wannier Hamiltonian in an interpolated fine **k**-mesh ($256 \times 256 \times 256$). The OHC is computed by the Kubo formula (details in Ref.[31]),

$$\sigma_{\alpha\beta}^{L_\gamma} = \frac{e\hbar}{2} \int \frac{d^3 k}{(2\pi)^3} \sum_{nn'} (f_{n\mathbf{k}} - f_{n'\mathbf{k}}) \frac{\text{Im}\left[\langle \psi_{n\mathbf{k}} | (v_\alpha L_\gamma + L_\gamma v_\alpha) | \psi_{n'\mathbf{k}} \rangle \langle \psi_{n'\mathbf{k}} | v_\beta | \psi_{n\mathbf{k}} \rangle \right]}{(E_{n\mathbf{k}} - E_{n'\mathbf{k}})^2 + \Gamma^2},$$

where $e > 0$ is the unit charge, $\hbar$ is the reduced Planck constant, $n$ and $n'$ are band indices, $\psi_{n\mathbf{k}}$ is a Bloch state with the energy eigenvalue $E_{n\mathbf{k}}$, $v_\alpha$ and $v_\beta$ are the velocity operators along $\alpha$ and $\beta$ directions, respectively, and $L_\gamma$ is the $\gamma$ component of the orbital angular momentum operator. For the calculation of the SHC, $L_\gamma$ is replaced by the spin operator $S_\gamma$. We introduce a phenomenological broadening by constant $\Gamma$, which we set 25 meV. Detailed results of the OHC and SHC calculated as a function of the Fermi energy is shown in Supplemental Note 4, Fig. S8.

# Symmetry-controlled orbital Hall effect in $IrO_2$

Supplemental Information

**Supplemental Note 1: Characterization of $IrO_2$ thin films**

High resolution x-ray diffraction (HR-XRD) was done for $IrO_2$ (001), (100), and (111) seen in Fig. S1 a, c, and e indicating a high epitaxial growth was achieved for the 3 orientations. Additionally, reciprocal space mapping (RSM) was performed for the 3 orientations seen in Fig. S2 indicating slight relaxation in the (001) orientation, anisotropic strain in the (100) orientation where $IrO_2$ grew fully coherent along the [010] direction and partially relaxed along the [001] direction, and (111) grew fully coherent along both in-plane crystallographic directions. Scanning transmission microscopy (STEM) was also done seen in Fig. S1 b, d, and f indicating sharp interfaces between $IrO_2$/$TiO_2$ and $IrO_2$/Py.

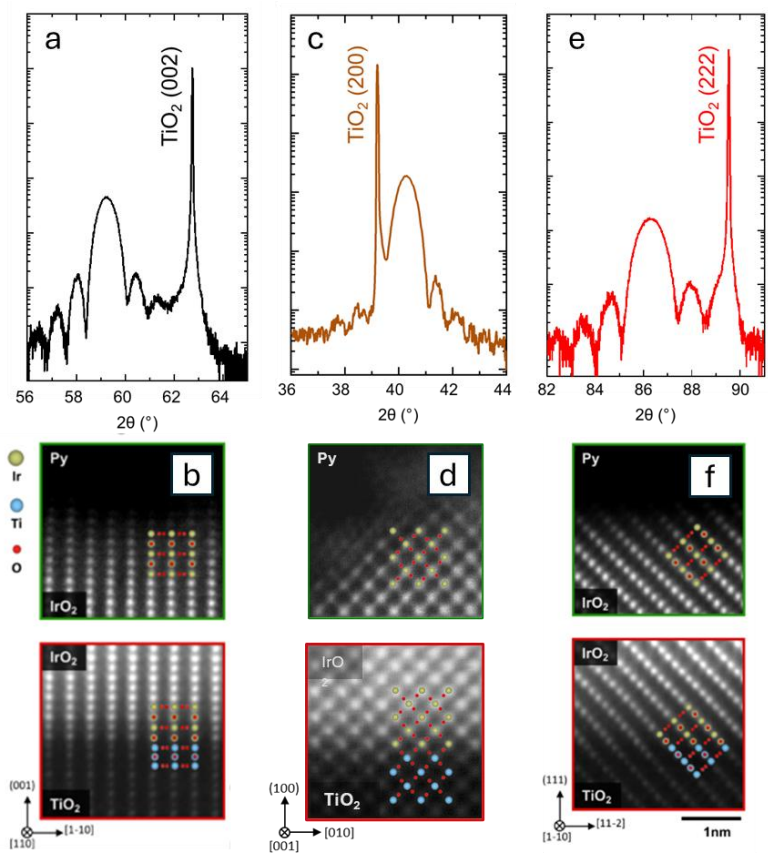

Figure S1: a, HR-XRD of the out of plane (002) peak and b, STEM of the interface between $IrO_2$ and Py and $TiO_2$ and $IrO_2$ with [1-10] zone axis. c, HR-XRD of the out of plane (200) peak and d, STEM of the interface between $IrO_2$ and Py and $TiO_2$ and $IrO_2$ with [010] zone axis. And e, HR-XRD of the out of plane (222) peak and f, STEM of the interface between $IrO_2$ and Py and $TiO_2$ and $IrO_2$ with [11-2] zone axis.

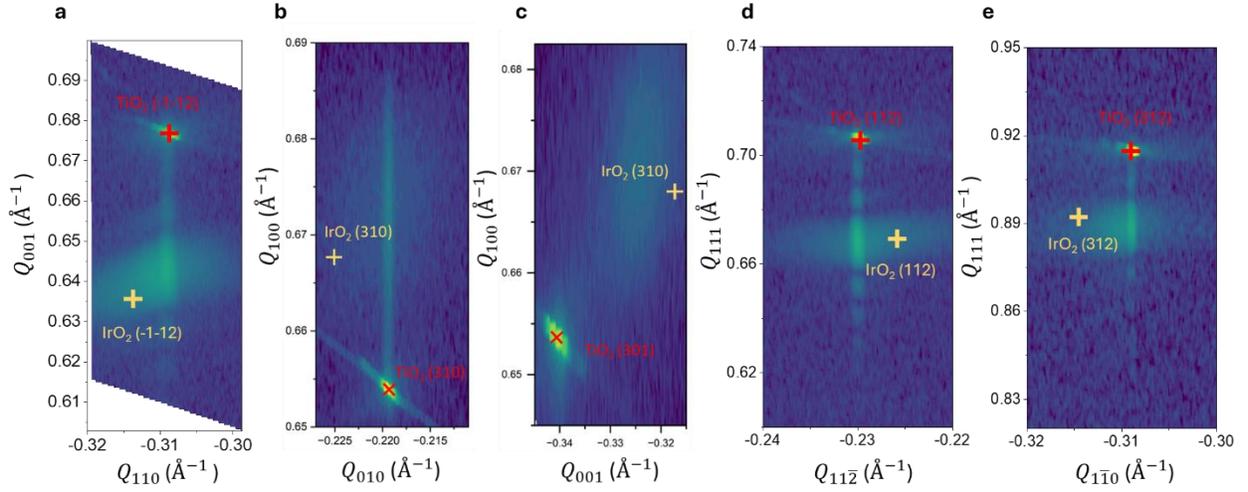

Figure S2: a, Reciprocal space map (RSM) of the (001) orientation of the (-1-12) peak, b and c RSMs of the (100) orientation, and d and e RSMs of the (111) orientation.

**Supplemental Note 2: ST-FMR lineshape analysis**

Measuring spin orbit torques can be done by detecting the magnetic dynamics of an adjacent ferromagnetic as spins are pumped into it. The magnetic dynamics under applied magnetic fields in the presence of damping-like spin orbit torques can be modelled using the Landau-Lifshitz-Gilbert-Slonczewski equation:

$$\frac{dM}{dt} = -\gamma M \times H_{eff} + \alpha M \times \frac{dM}{dt} + \tau_{DL} M \times (\sigma \times M) \qquad (1)$$

Where M is the magnetization, $\gamma$ is the gyromagnetic ratio, $H_{eff}$ is the applied magnetic field, $\alpha$ is the Gilbert damping coefficient, $\tau_{DL}$ is the damping-like torque coefficient, and $\sigma$ is the polarization direction of the spin current. There are several techniques to experimentally determine this, the most common being spin torque ferromagnetic resonance measurements.

During the ST-FMR measurements, a microwave current is applied at a fixed frequency and fixed power while sweeping an in-plane magnetic field through the ferromagnetic resonance conditions. The microwave current was modulated at a fixed frequency which can be detected using a lock-in amplifier by measuring the DC mixing voltage across the device at the same modulated frequency. The mixing voltage was fitted vs applied field to extract the symmetric and antisymmetric Lorentzian components. Conventionally, torques generated from Oersted fields acting on the magnetization are in the form of $\boldsymbol{m} \times \boldsymbol{y}$ where $\boldsymbol{y}$ is an in-plane direction perpendicular to the charge current direction $\boldsymbol{x}$ resulting in a field-like torque ($\tau_\perp$). In addition to the out-of-plane torques, spin currents resulting from the SHE (or other spin generating effects) can also generate torques in the form of $\mathbf{m} \times (\mathbf{m} \times \mathbf{y})$ resulting in a damping-like torque ($\tau_\parallel$). The out-of-plane ($\tau_\perp$) and the in-plane ($\tau_\parallel$) torques are proportional to the mixing voltage $V_{mix}$ as the ferromagnetic layer goes through its resonance condition which can be fitted as a sum of a symmetric and an antisymmetric Lorentzian:

$$V_{mix,S} = -\frac{I_{rf}}{2}\left(\frac{dR}{d\varphi}\right)\frac{1}{\alpha(2\mu_0 H_{FMR}+\mu_0 M_{eff})}\tau_\parallel \quad (2)$$

$$V_{mix,A} = -\frac{I_{rf}}{2}\left(\frac{dR}{d\varphi}\right)\frac{\sqrt{1+M_{eff}/H_{FMR}}}{\alpha(2\mu_0 H_{FMR}+\mu_0 M_{eff})}\tau_\perp \quad (3)$$

Where R is the resistance of the device, $\varphi$ is the magnetization angle with respect to the applied current, $\alpha$ is the Gilbert damping coefficient, $\mu_0 H_{FMR}$ is the resonance field, and $\mu_0 M_{eff}$ is the effective magnetization. The effective magnetization of Py can be obtained using Kittel's equation $f = \frac{\gamma}{2\pi}\sqrt{(H_{FMR} + H_K)(H_{FMR} + H_K + M_{eff})}$ where $\gamma$ is the gyromagnetic ratio and $H_K$ is the in-plane anisotropy field. The Gilbert damping coefficient $\alpha$ is obtained by fitting the linear relationship between the linewidth ($w$) and the frequency $w = w_0 + \left(\frac{2\pi}{\gamma}\right) * f$. Fig. S3 a, c, and e shows the frequency dependent ST-FMR for IrO$_2$ (001)/Ni, IrO$_2$ (001)/Pt/Py/Pt, and Pt/Py/Pt control sample, respectively, and Fig. S3 b, d, and f shows the angular ST-FMR for IrO$_2$ (001)/Ni, IrO$_2$ (001)/Pt/Py/Pt, and Pt/Py/Pt control sample, respectively.

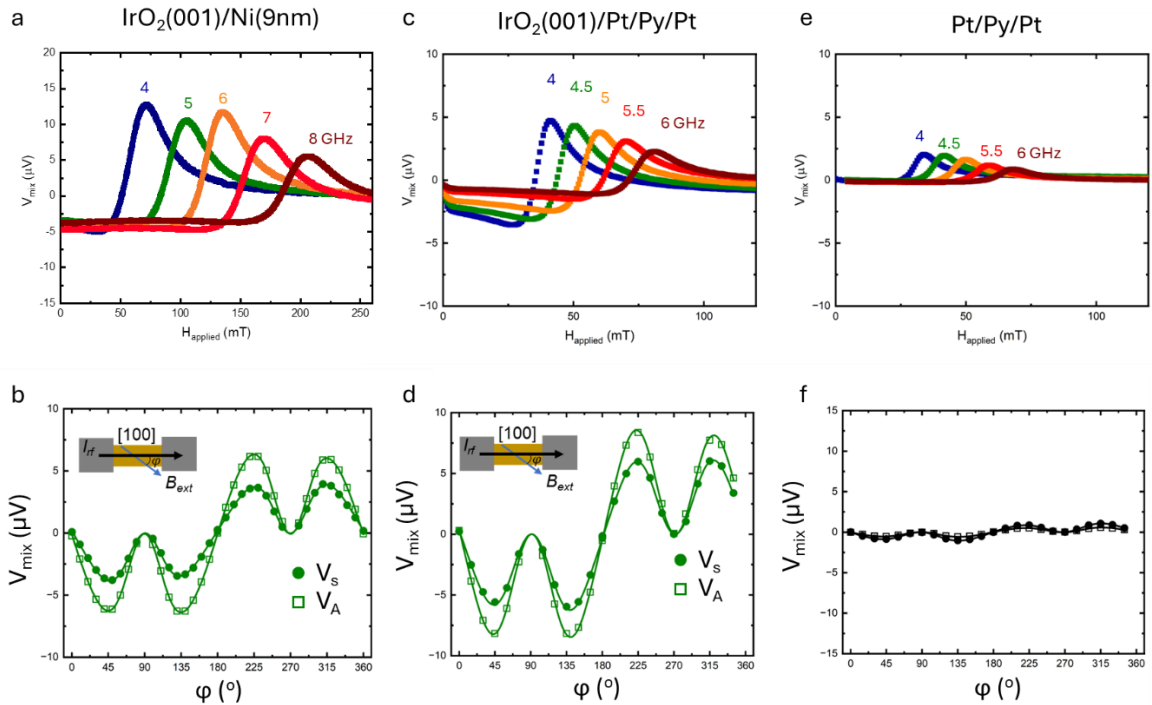

Figure S3: a, ST-FMR lineshape analysis vs frequency for IrO$_2$/Ni. b, Angular ST-FMR IrO$_2$/Ni. c, ST-FMR lineshape analysis vs frequency for IrO$_2$/Pt/Py/Pt. d, Angular ST-FMR IrO$_2$/Pt/Py/Pt. e, ST-FMR lineshape analysis vs frequency for Pt/Py/Pt control sample. And f, Angular ST-FMR Pt/Py/Pt control sample.

Calibration of the microwave current ($I_{rf}$) can be determined by measuring the resistance change due to Joule heating across the Hall bar device while varying the microwave power. We can then find $I_{rf} = \sqrt{2}I_{dc}$ due to the joule heating relationship between AC and DC current. Fig. S4 a and c show the Joule heating for varying DC current and RF power for IrO$_2$ (001)/Ni (6nm) and IrO$_2$ (001)/Pt/Py/Py, respectively. Using RF calibrated current determination, we calculate the expected Oersted field for the 2 samples and compare with the antisymmetric voltage signal measured from angular ST-FMR, indicating that the antisymmetric voltage is primarily coming from the Oersted field seen in Fig. S4 b and d.

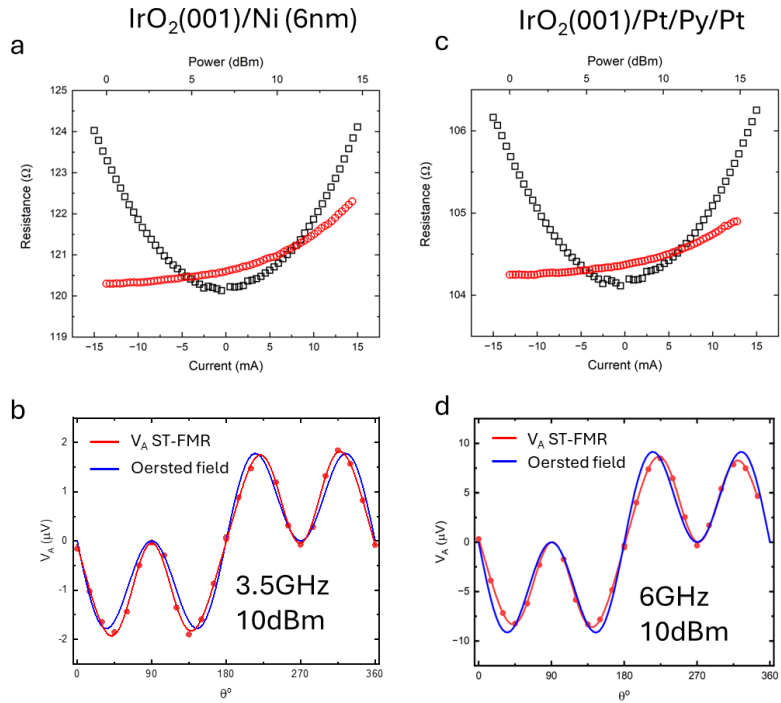

Figure S4: a, RF current calibration determined by measuring the change in resistance due to Joule heating by varying the applied DC current and RF power for IrO$_2$ (001)/Ni. b, Angular ST-FMR for IrO$_2$ (001)/Ni (red) with Oersted contributions (blue) determined using RF current calibration. c, RF current calibration determined by measuring the change in resistance due to Joule heating by varying the applied DC current and RF power for IrO$_2$ (001)/Pt/Py/Pt. d, Angular ST-FMR for IrO$_2$ (001)/ Pt/Py/Pt (red) with Oersted contributions (blue) determined using RF current calibration.

To quantitatively determine all torque contributions, ST-FMR measurements were performed as a function of applied in-plane magnetic field angle ($\varphi$). In conventional heavy-metal/ferromagnetic bilayers the AMR of the ferromagnetic layer ($\frac{dR}{d\varphi}$) has an angular dependance proportional to $\sin(2\varphi)$ and the out-of-plane ($\tau_\perp$) and the in-plane ($\tau_\parallel$) torques are proportional to $\cos(\varphi)$ resulting in the $V_{mix,S}$ and $V_{mix,A}$ being in the form of $\sin(2\varphi)\cos(\varphi)$. Additional unconventional torques that have different spin polarization directions, however, can contribute to the angular dependance resulting in a more general form of the angular dependance:

$$\tau_\parallel = \tau_{x,AD} \sin(\varphi) + \tau_{y,AD} \cos(\varphi) + \tau_{z,FL} \quad (4)$$
$$\tau_\perp = \tau_{x,FL} \sin(\varphi) + \tau_{y,FL} \cos(\varphi) + \tau_{z,AD} \quad (5)$$

$V_{mix,S}$ and $V_{mix,A}$ can then be expressed in the form of $\sin(2\varphi)(\tau_{x,AD} \sin(\varphi) + \tau_{y,AD} \cos(\varphi) + \tau_{z,FL})$ and $\sin(2\varphi)(\tau_{x,FL} \sin(\varphi) + \tau_{y,FL} \cos(\varphi) + \tau_{z,AD})$, respectively.

Additional angular ST-FMR measurements for increasing Ni thickness can be seen in Fig. S5. From these measurements, it is clear that as the Ni thickness increases the ratio between the symmetric and antisymmetric voltages tends to decrease up to 10nm of Ni and then begins to saturate. The effective magnetization for the various Ni thicknesses can be seen in Fig. S6 where we see a slight increase in magnetization of Ni with increasing thickness.

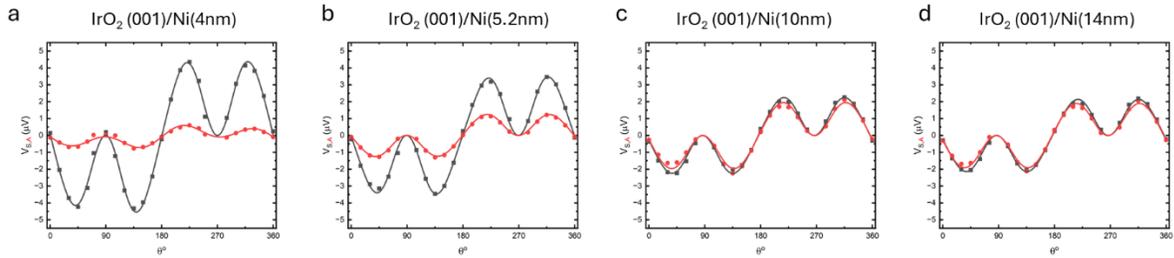

Figure S5: a-d Angular ST-FMR for $IrO_2$ (001)/Ni for increasing Ni thickness for $V_s$ (black) and $V_s$ (red).

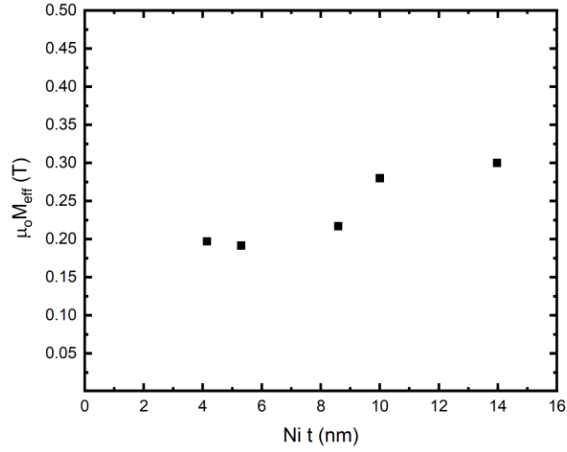

Figure S5: Effective magnetization of Ni for various Ni thickness determined from the ST-FMR lineshape analysis.

Supplemental Note 3: Orbital-spin conversion through Pt layer

Another approach to probe the OHC is using Pt bilayers which has been proposed and backed with experimental evidence to convert orbital currents to spin currents which can then be detected as a torque on an adjacent ferromagnet material. Although Pt itself can generate a large spin Hall effect, by growing a Pt/Py/Pt heterostructure where both Pt layers have the same thickness we can effectively cancel out all contributions from Pt. Based on other reports, we chose a thickness of ~1.5nm of Pt which should allow for maximum conversion[1]. The finite symmetric voltage signal we found in the Pt/Py/Pt control sample (see Supplemental Note 2), is likely due to RF current shunting into the $TiO_2$ substrate due to high dielectric loss which is a known problem in other oxide substrates[2]. To minimize signals coming from the substrate, we measure DC-tuned ST-FMR which applies a constant dc current in addition to the rf current during ST-FMR measurements, which can be used to isolate the true SHC of a material as the rf current shunting should not be affected by the applied dc current[2–4]. The linear relationship between the linewidth of the mixing voltage signal during ST-FMR measurements and the DC current can be used to determine the SHC which is proportional to $\frac{\Delta\alpha_{eff}}{\Delta J_c}$ where the sign of the slope indicates the sign of the SHC. During the DC-tuned measurements, a DC bias was applied at currents ranging between -2 to 2 mA in addition to a fixed rf current by using a bias tee. The spin Hall angle can be determined using the following equation:

$$\theta_{DL} = \frac{2e}{\hbar} \left( \frac{(H_{FMR} + \frac{M_{eff}}{2})\mu_0 M_s t_{Py}}{\sin(\varphi)} \right) \frac{\Delta\alpha_{eff}}{\Delta J_c}$$

Where $t_{Py}$ is the thickness of permalloy and $\frac{\Delta\alpha_{eff}}{\Delta J_c}$ is the linear slope of effective damping coefficient, determined from the linear relationship between the linewidth ($w$) and the frequency $w = w_0 + \left(\frac{2\pi}{\gamma}\right) * f$, vs the charge current going through the IrO$_2$ layer determined using parallel resistor model. Finally, we can obtain the spin Hall conductivity from $\sigma^i_{jk} = \frac{\theta_i}{\rho_{IrO2}}\frac{\hbar}{2e}$. In the Pt/Py/Pt control sample, we find no change in the damping coefficient vs varying DC current seen in Fig. S6b. We then measured the DC-tuned ST-FMR for IrO$_2$ (001) and (100) with Pt/Py/Pt. For the (001) orientation seen in Fig. S6a, we find the SHC to be 188 $\pm$ 16 ($\frac{\hbar}{e}$ ($\Omega$ cm)$^{-1}$) compared to IrO$_2$ (001)/Py of 247 $\pm$ 14 ($\frac{\hbar}{e}$ ($\Omega$ cm)$^{-1}$). For the (100) orientation with charge current along the [001] direction seen in Fig. S6b, we find the SHC to be 310 $\pm$ 21 ($\frac{\hbar}{e}$ ($\Omega$ cm)$^{-1}$) compared to IrO$_2$ (100)/Py of 190 $\pm$ 50 ($\frac{\hbar}{e}$ ($\Omega$ cm)$^{-1}$). Lastly, for the (100) orientation with charge current along the [010] direction seen in Fig. S6d, we find the SHC to be -171 $\pm$ 8 ($\frac{\hbar}{e}$ ($\Omega$ cm)$^{-1}$) compared to IrO$_2$ (100)/Py of -34 $\pm$ 14 ($\frac{\hbar}{e}$ ($\Omega$ cm)$^{-1}$). In Table S1, we summarize the results from IrO$_2$/Py which should not be sensitive to orbital currents, to the IrO$_2$/Pt/Py/Pt samples for both angular ST-FMR and DC-tuned ST-FMR.

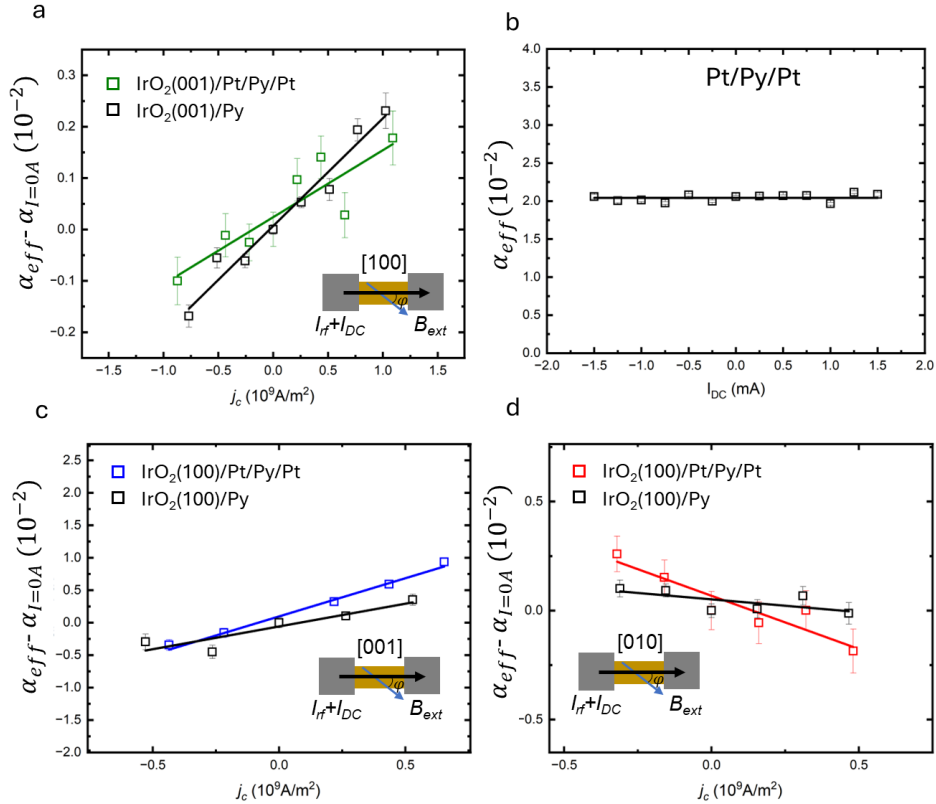

Figure S6: a DC-tuned ST-FMR for IrO$_2$(001)/Pt/Py/Pt and IrO$_2$(001)/Py, b DC-tuned ST-FMR for IrO$_2$(100)/Pt/Py/Pt and IrO$_2$(100)/Py with charge current along the [001] direction, c DC-

tuned ST-FMR for IrO$_2$(100)/Pt/Py/Pt and IrO$_2$(100)/Py with charge current along the [010] direction, and d, DC-tuned ST-FMR for Pt/Py/Pt control sample.

Table S1: Spin Hall angle $\theta^y$ and spin Hall conductivity $\sigma^y$ ($\frac{\hbar}{e}$ ($\Omega$ cm)$^{-1}$) for IrO$_2$/Py and IrO$_2$/Pt/Py/Pt bilayers for angular ST-FMR and DC-tuned ST-FMR measurements.

| Sample | $\theta^y_{DC-tuned}$ | $\sigma^y_{DC-tuned}$ |
|---|---|---|
| IrO$_2$(001)/Py | 0.035 ± 0.003 | 247 ± 21 |
| IrO$_2$(001)/Pt/Py/Pt | 0.036 ± 0.002 | 188 ± 16 |
| IrO$_2$(100)/Py [00$\bar{1}$] | 0.06 ± 0.015 | 190 ± 50 |
| IrO$_2$(100)/Pt/Py/Pt [00$\bar{1}$] | 0.068 ± 0.003 | 310 ± 21 |
| IrO$_2$(100)/Py [010] | -0.03 ± 0.006 | -34 ± 14 |
| IrO$_2$(100)/Pt/Py/Pt [010] | -0.057 ± 0.002 | -171 ± 8 |

**Supplemental Note 4: First-principles calculations**

We check the consistency of the band structures obtained by the full-potential linearly augmented plane wave (FLAPW) and Wannier function methods. Figure S7 shows good agreement between the band structures calculated by the two methods.

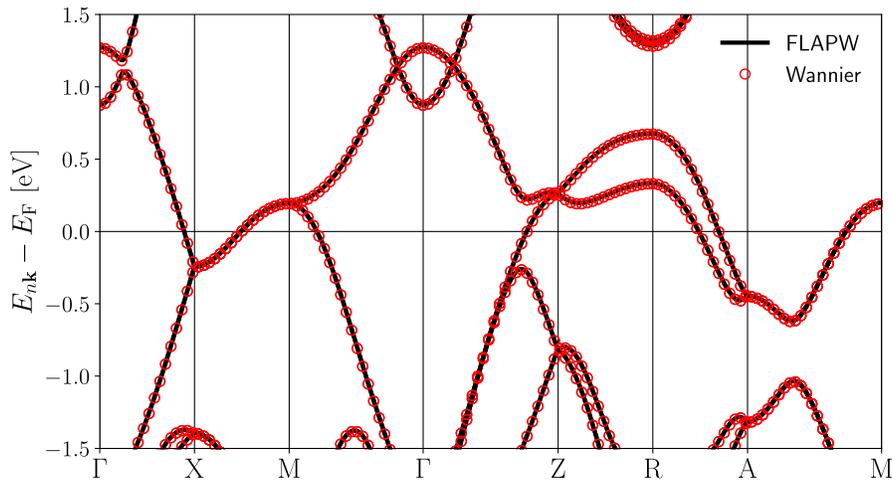

Figure S7: Comparison of the band structures obtained by the FLAPW and Wannier function methods, which show good agreement.

Details of the results from the first-principles calculations for the OHC and SHC in $IrO_2$ is shown in Fig. S8. Three independent tensor components of the OHC and SHC are plotted as a function of the Fermi energy, which is varied within the frozen band approximation for a fixed potential.

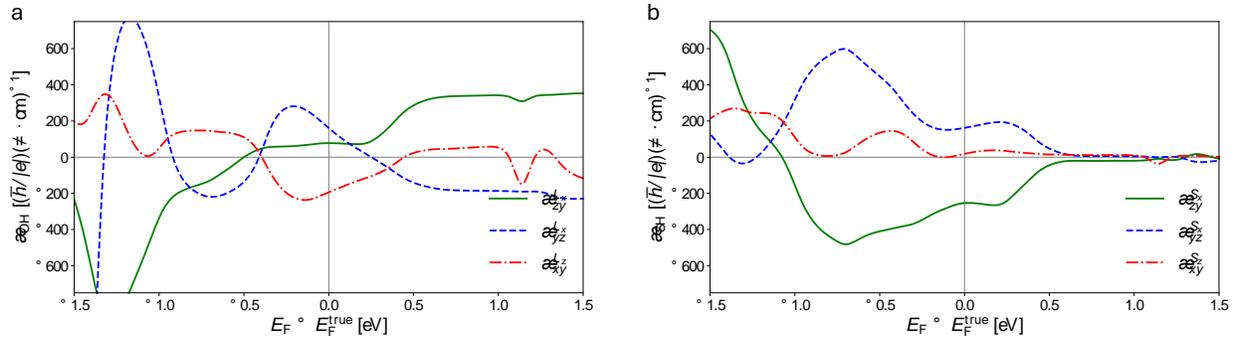

Figure S8: SHC (a) and OHC (b) for the 3 non-zero components in the (001) basis as a function of the Fermi energy.